\PassOptionsToPackage{numbers}{natbib}
\documentclass{article}
\usepackage[preprint]{neurips_2025}
\usepackage[utf8]{inputenc} 
\usepackage[T1]{fontenc}    
\usepackage{hyperref}       
\usepackage{url}            
\usepackage{booktabs}       
\usepackage{amsfonts}       
\usepackage{nicefrac}       
\usepackage{microtype}      
\usepackage{xcolor}         
\usepackage{amsmath}        
\usepackage{array}          
\usepackage{longtable}      
\usepackage{enumitem}       
\usepackage{graphicx}       
\title{A Penny for Your Thoughts: Decoding Speech from Inexpensive Brain Signals}

\author{
  Quentin Auster\thanks{Equal contribution.}\textsuperscript{1}, 
  Kateryna Shapovalenko\footnotemark[1]\textsuperscript{1}, 
  Chuang Ma\footnotemark[1]\textsuperscript{1}, 
  Demaio Sun\footnotemark[1]\textsuperscript{1} \\
  \textsuperscript{1}Carnegie Mellon University, Pittsburgh, PA 15213 \\
  \texttt{\{qja, kshapova, chuangm, cleons\}@alumni.cmu.edu}
}

\begin{document}
\maketitle
\begin{abstract}
  We explore whether neural networks can decode brain activity into speech by mapping EEG recordings to audio representations. Using EEG data recorded as subjects listened to natural speech, we train a model with a contrastive CLIP loss to align EEG-derived embeddings with embeddings from a pre-trained transformer-based speech model. Building on the state-of-the-art EEG decoder from Meta, we introduce three architectural modifications: (i) \textbf{subject-specific attention layers} (+0.15\% WER improvement), (ii) \textbf{personalized spatial attention} (+0.45\%), and (iii) a \textbf{dual-path RNN with attention} (-1.87\%). Two of the three modifications improved performance, highlighting the promise of personalized architectures for brain-to-speech decoding and applications in brain-computer interfaces.
\end{abstract}


\section{Introduction}

\begin{figure}[ht]
    \centering
    \includegraphics[width=0.7\textwidth]{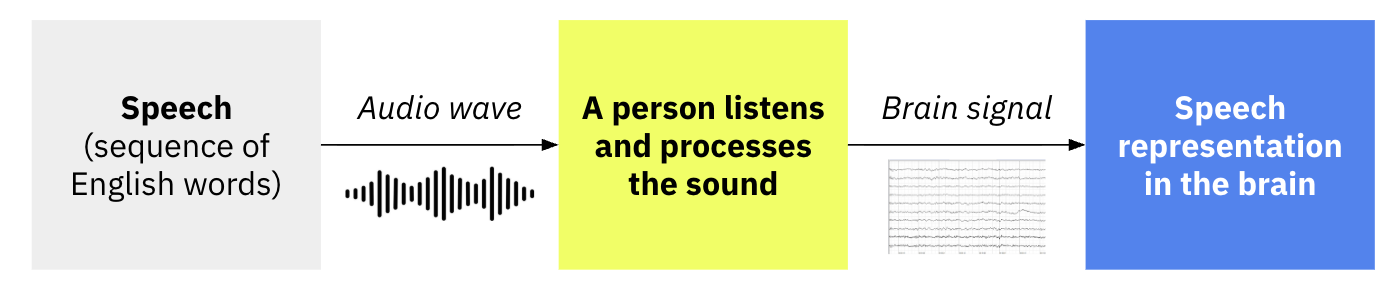}
    \caption{From Sound to Brain Representation.}
    \label{fig:motivation}
\end{figure}

Human eardrums respond to minute and rapid changes in air pressure, a seemingly simple mechanism that underlies the rich auditory experiences we perceive. Despite this simplicity, the auditory system can not only capture sounds but also distinguish between multiple simultaneous sources through frequency tuning in the cochlea \cite{oxenham2018}. From a single incoming sound wave, humans are able to separate and selectively attend to different acoustic streams.

Speech cues are encoded at the subcortical level and further shaped by both short- and long-term auditory experiences \cite{holt2022}. These cues are distributed across cortical regions that specialize in extracting specific features of speech. Lower-level sound features are progressively transformed into higher-level speech representations that support perception and comprehension \cite{holt2022}. Core auditory areas represent rich acoustic features, which are then relayed to non-core auditory regions, such as the lateral superior temporal gyrus, for mapping onto phonemic and linguistic structures.

Given this structured flow from sound to brain representations, we hypothesize that the inverse mapping from brain signals back to speech can be approximated using deep neural networks. Specifically, we investigate whether it is possible to predict sequences of English words that a subject has just heard from EEG recordings of their brain activity (Figure~\ref{fig:motivation}). While the current state-of-the-art for neural speech decoding leverages Magnetoencephalography (MEG), EEG offers a significantly more accessible and cost-effective modality. Accurate decoding of speech from EEG could greatly enhance accessibility for speech-impaired individuals and improve brain-computer interface technologies.

In this paper, we use EEG data collected while subjects listened to a chapter from Alice in Wonderland \cite{brennan_2019}. Building on the recent work of D\'efossez et al.\ (Meta) \cite{defossez2023decoding}, we propose architectural modifications that extend their subject-specific design to improve decoding performance. Our model integrates subject-specific spatial attention, subject-level attention layers, convolutional blocks, and recurrent components. Evaluated on a subset of the original dataset, our approach demonstrates promising improvements in word error rate (WER) over Meta's baseline, suggesting that architectural personalization is a valuable direction for brain-to-speech decoding.


\section{Literature Review}

The field of speech decoding from brain activity is a rapidly evolving area of research, characterized by its challenges and the variety of approaches employed. We can gain insights into this field by focusing on two key aspects: (i) the type of brain signal recordings used and (ii) the models applied for speech decoding from the brain. A structured overview of the relationships between brain signal types, their characteristics, and model architectures is provided in Table~\ref{tab:brain_signals}.

\paragraph{Types of Brain Signal Recordings} Research in this field predominantly uses either non-invasive methods like Electroencephalography (EEG), Magnetoencephalography (MEG), and Functional Magnetic Resonance Imaging  (fMRI), or invasive methods such as Electrocorticography (ECoG). Non-invasive methods, while more accessible and less risky, often provide less rich signal data compared to invasive methods. Invasive methods, despite their higher risk and complexity, tend to yield more accurate results in speech decoding due to richer data quality. In our research, we concentrate on non-invasive methods (specifically, EEG), each with unique characteristics. EEG offers excellent temporal resolution but falls short in spatial resolution, limiting its effectiveness in detailed spatial analysis of brain activity. In contrast, MEG provides high precision in both temporal and spatial aspects, generally leading to better decoding results. For instance, D\'efossez et al. \cite{defossez2023decoding} observed significantly improved performance using MEG data over EEG. fMRI, with its superior spatial resolution but limited temporal resolution, struggles in capturing the precise timing of events. The scanner noise in fMRI can also interfere with auditory responses. However, its high spatial resolution has been instrumental in achieving notable accuracy in speech decoding. This underscores the importance of spatial resolution in decoding speech perception, a factor that should be carefully considered when working with EEG signals.

\paragraph{Models for Speech Decoding from Brain} The decoding of speech from brain activity has seen various approaches using combinations of convolutional, recurrent, and sequence-to-sequence architectures. D\'efossez et al. (2023) \cite{defossez2023decoding} utilized a transformative approach, employing a pre-trained transformer-based speech model (\texttt{wav2vec2} \cite{baevski2020}) to process audio recordings. They combined this with a convolutional neural network featuring a subject-specific layer for extracting representations from MEG and EEG recordings. The use of a contrastive loss, CLIP Loss \cite{radford2021}, allowed them to train a zero-shot decoding classifier effectively. Tang et al. (2023) \cite{tang2023} introduced an innovative brain signal decoder designed to reconstruct continuous language from fMRI across various tasks. Their model included a beam search decoder, generating candidate sequences of words, alongside a GPT-based language model, demonstrating the potential of integrating advanced language processing models in speech decoding. Zhang et al. (2018) \cite{zhang2018} explored a combination of convolutional and recurrent neural networks to decode brain activities from motion imagery EEG recordings. They also incorporated an autoencoder layer to filter out background activity, highlighting the importance of noise reduction in enhancing decoding accuracy.

\newpage
\setlength{\heavyrulewidth}{1.5pt}
\setlength{\abovetopsep}{4pt}
\begin{longtable}[t]{p{0.15\linewidth}p{0.35\linewidth}p{0.4\linewidth}}
\caption{Brain Signal Types and Models of Speech Decoding\label{tab:brain_signals}} \\
\toprule
\textbf{Brain Signals} & \textbf{Characteristics of Brain Signals} & \textbf{Models for Speech Decoding} \\
\midrule
\endfirsthead

\caption[]{Brain Signal Types and Models of Speech Decoding (continued)} \\
\toprule
\textbf{Brain Signals} & \textbf{Characteristics of Brain Signals} & \textbf{Models for Speech Decoding} \\
\midrule
\endhead

EEG 
& 
\begin{itemize}[leftmargin=*,noitemsep,topsep=0pt,partopsep=0pt] 
    \item Non-invasive; measures electrical activity of neurons.
    \item High temporal, low spatial resolution.
    \item Noise: (1) Environmental (power lines, electronics), (2) Physiological (blinks, heartbeat), (3) Electrode contact.
\end{itemize}
& 
\begin{itemize}[leftmargin=*,noitemsep,topsep=0pt,partopsep=0pt]
    \item Défosséz et al. (2023) \cite{defossez2023decoding}:
    \begin{itemize}[noitemsep]
        \item Used pre-trained transformer-based model (\texttt{wav2vec2}) on audio.
        \item EEG processed with CNN + subject-specific layer.
        \item Trained with contrastive CLIP loss \cite{radford2021} for zero-shot classification.
        \item Achieved 25.75\% vocabulary-specific accuracy.
    \end{itemize}
    \item Zhang et al. (2018) \cite{zhang2018}:
    \begin{itemize}[noitemsep]
        \item Hybrid CNN-RNN (LSTM) to extract features from motor imagery EEG.
        \item Autoencoder to filter background noise.
        \item Achieved 95.53\% accuracy of final classification with XGBoost.
    \end{itemize}
\end{itemize}
\\
\midrule

MEG 
& 
\begin{itemize}[leftmargin=*,noitemsep,topsep=0pt,partopsep=0pt]
    \item Non-invasive; measures magnetic fields from neural activity.
    \item High temporal and spatial resolution.
    \item Noise: (1) Magnetic interference (requires shielding), (2) Subject movement.
\end{itemize}
& 
\begin{itemize}[leftmargin=*,noitemsep,topsep=0pt,partopsep=0pt]
    \item Défosséz et al. (2023) \cite{defossez2023decoding}:
    \begin{itemize}[noitemsep]
        \item Same architecture as EEG setup.
        \item Achieved 66.69\% vocabulary-specific accuracy.
    \end{itemize}
\end{itemize}
\\
\midrule

fMRI 
& 
\begin{itemize}[leftmargin=*,noitemsep,topsep=0pt,partopsep=0pt]
    \item Non-invasive; measures blood-oxygen-level changes.
    \item High spatial, low temporal resolution.
    \item Noise: (1) Scanner noise, (2) Physiological noise, (3) Motion artifacts.
\end{itemize}
& 
\begin{itemize}[leftmargin=*,noitemsep,topsep=0pt,partopsep=0pt]
    \item Tang et al. (2023) \cite{tang2023}:
    \begin{itemize}[noitemsep]
        \item Reconstructed continuous language using beam search decoder.
        \item Integrated GPT-based language model.
    \end{itemize}
    \item Caucheteux et al. (2022) \cite{caucheteux2022}:
    \begin{itemize}[noitemsep]
        \item Used GPT-2 model to predict semantic comprehension from fMRI.
    \end{itemize}
\end{itemize}
\\
\bottomrule
\end{longtable}


\section{Dataset}

We chose to work with EEG data as it is non-invasive and inexpensive to record compared to other brain signal recording methods such as MEG or fMRI. We rely on data from Brennan and Hale (2019) \cite{brennan_2019}, which is also used by Meta (our baseline reference). This dataset contains EEG data collected using 62 sensors from 33 subjects, totaling approximately 6.7 hours of recordings. 

\paragraph{EEG Data} Brennan and Hale recorded EEG data while participants listened to spoken prose from Chapter One of Alice in Wonderland. EEG was recorded for 49 participants. However, the recordings of 16 participants were not used due to noise in the data, poor results on a listening comprehension test after recording, or both, leaving recordings from 33 subjects \cite{brennan_2019}. EEG was recorded from 62 sensors, or channels, including "VEOG" (Vertical Electrooculogram) and "AUD" (Auditory) channels. VEOG is used to measure vertical eye movements, which can create artifacts in EEG data \cite{jiang2019}. The EEG data was recorded at a sampling rate of 500 Hz (in the data, timestamps increment by 0.002 seconds) in BrainVision Core Data format (.vhdr, .vmark, and .eeg files provide the metadata, events collected, and EEG data with additional signals, respectively).

\paragraph{Audio Data} Audio data recording the reading aloud of Chapter One of Alice in Wonderland is provided in 12 wav files (segments). The paper's authors also provide a table with each word spoken in the chapter, along with its starting and ending time in the corresponding segment.

\begin{figure}[ht]
    \centering
    \includegraphics[width=0.8\textwidth]{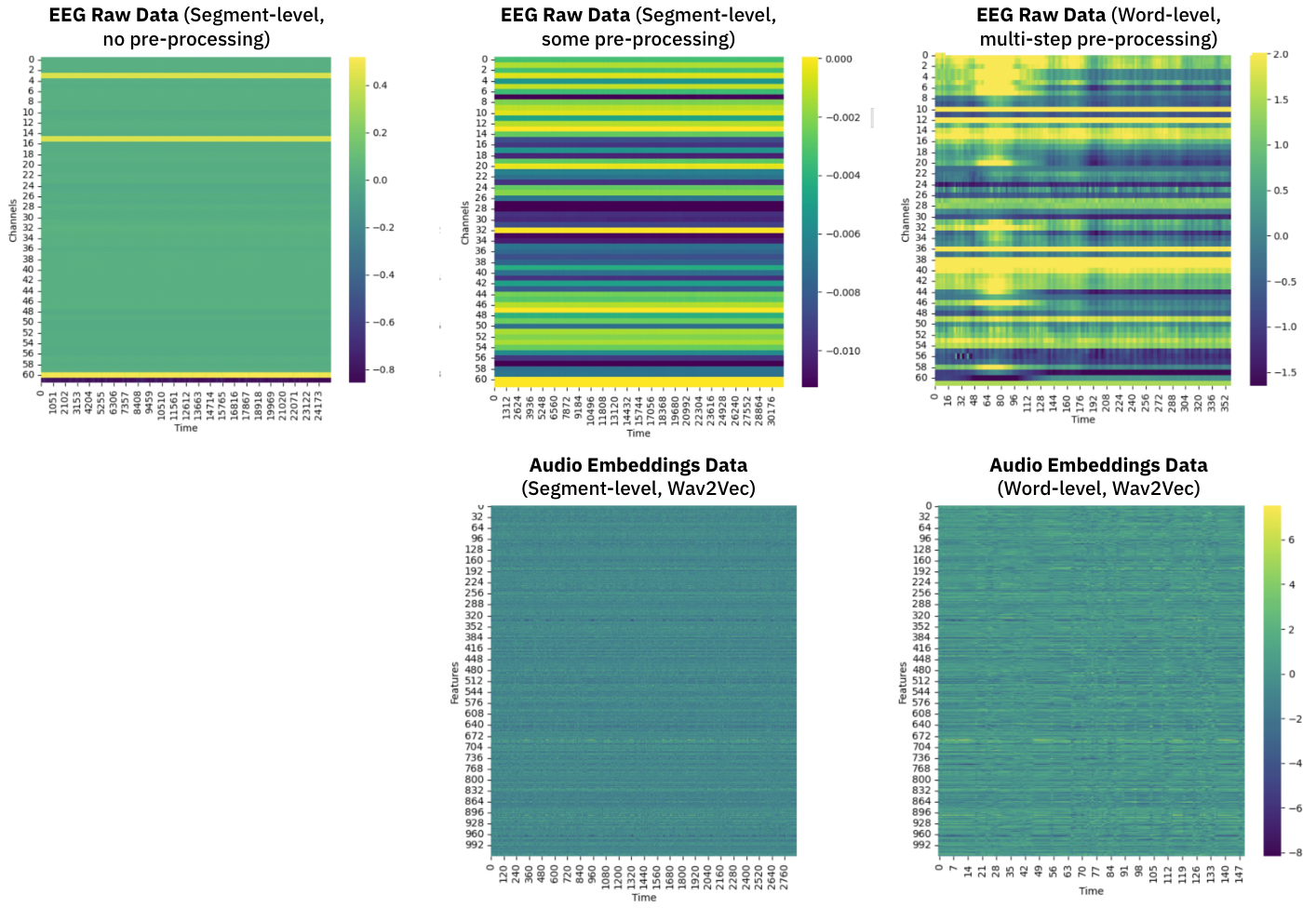}
    \caption{Pre-Processing of EEG and Audio Data}
    \label{fig:preprocessing}
\end{figure}

\paragraph{Data Pre-processing} We follow Meta's approach for EEG and audio pre-processing. For EEG data, we begin by loading the raw data, then apply baseline correction for signal stability and robust scaling for variance consistency. We further refine the data by clipping outliers below the 5th percentile and above the 95th percentile, and clamping values exceeding 20 standard deviations. Both EEG and audio data undergo standard normalization. We segment these data into three-second windows to focus on word-level decoding. While initially experimenting with segment-level data to capture broader neural patterns, we faced challenges in model convergence, indicating the importance of segment length in EEG-audio studies. We leave further experimentation with segment length as an area for future work. The results of this pre-processing pipeline is shown in Figure~\ref{fig:preprocessing}.

\section{Model Description}

Let \(X \in \mathbb{R}^{T \times C}\) be a segment of EEG recording with \(C\) channels and \(T\) time steps, and let \(Y \in \mathbb{R}^{T \times F}\) represent the corresponding audio segment (e.g., a Mel spectrogram or hidden representation from a pre-trained speech model), with \(F\) channels and \(T\) aligned time steps. The task of decoding brain signals into speech can be formulated as learning a function \(\mathbf{f}: \mathbb{R}^{C \times T} \rightarrow \mathbb{R}^{F \times T}\). Once the model \(\mathbf{f}\) maps the EEG input to the same space and temporal resolution as the audio, the resulting embeddings can be compared. For inference and validation, we follow Meta’s approach of passing hidden states from both our model and the pre-trained speech model through the final layer(s) of the latter.

\begin{figure}[ht]
    \centering
    \includegraphics[width=1\textwidth]{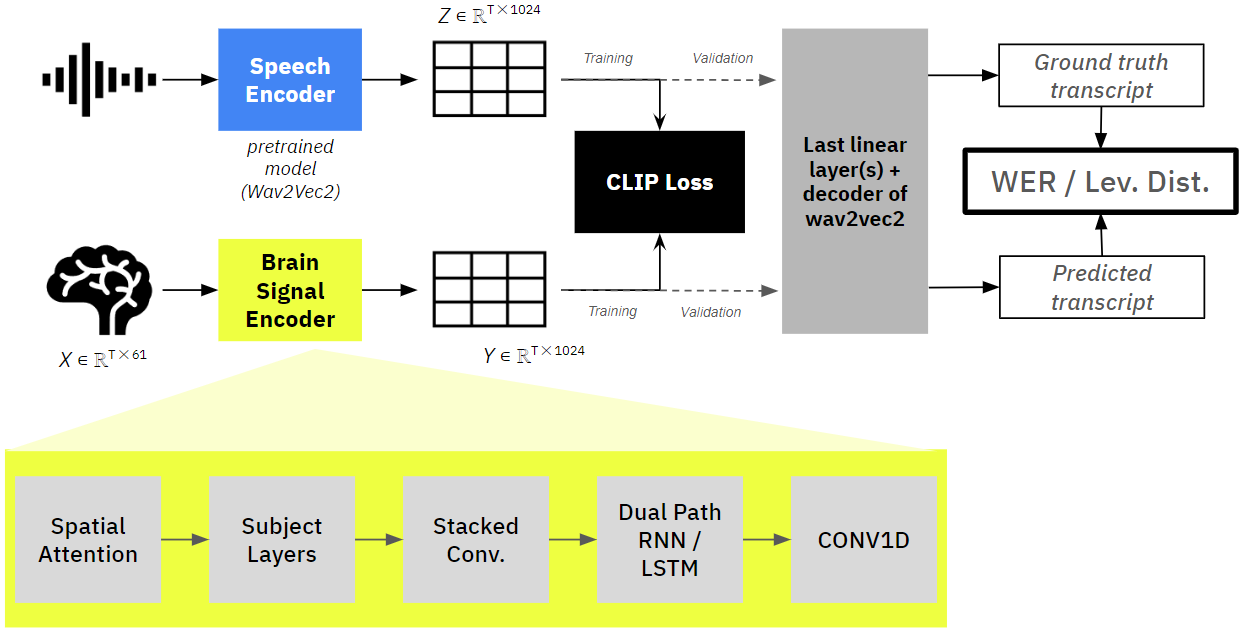}
    \caption{Model Architecture}
    \label{fig:model}
\end{figure}

After analyzing EEG properties and running extensive experiments, we identified several effective modifications to Meta’s architecture. We used Meta’s codebase as a foundation and focused on five key components: (1) Adaptive Spatial Attention, (2) Subject-Specific Attention, (3) Stacked Convolutions, (4) Dual Path RNN with Attention, and (5) Final Convolutions (Figure~\ref{fig:model}).

\paragraph{Adaptive Spatial Attention} Our model extends Meta’s by introducing a subject-specific spatial attention mechanism. As in Meta, we use \(\texttt{mne.channels.find\_layout}\) to project 3D sensor positions to 2D and normalize them to \([0,1]\). For each output channel, a learned function, parameterized in Fourier space, assigns values in \([0,1]^2\), and spatial dropout is applied to mitigate overfitting. While Meta used a shared attention map across all subjects, our \(\texttt{SubjectAttentionLayers}\) module assigns unique attention weights to each subject by using the \(\texttt{SubjectAttention}\) class. This enables personalized spatial filtering, accommodating inter-subject variability. We found this approach more aligned with individual brain activation patterns, improving over Meta’s design, where attention maps did not consistently focus on auditory regions (see Extended Data Figure 5 in \cite{defossez2023decoding}).

\paragraph{Subject-Specific Attention} We further introduce a subject-specific layer after spatial attention for each participant. This layer uses the \(\texttt{SubjectAttentionLayers}\) module, combining \(\texttt{SubjectLayers}\) and \(\texttt{SubjectAttention}\), to model unique patterns in each subject’s brain data. This improves the model's ability to adapt to inter-individual neural differences, offering a more accurate and personalized decoding.

\paragraph{Stacked Convolutions} Our model adopts a stack of convolutional blocks, implemented via the \(\texttt{ConvSequence}\) class. Each block, built using \(\texttt{ResidualBlock}\), contains two convolutional layers with residual connections and expands the output to \(D_2 = 320\) channels. Batch normalization ensures training stability. Convolution dilation follows the pattern \(2^{2k \bmod 5}\) and \(2^{2k+1 \bmod 5}\), enhancing the receptive field. A Gated Linear Unit (GLU) activation reduces the channel size by half. This setup mirrors Meta’s structure and ensures strong temporal modeling without modifying the convolution backbone.

\paragraph{Dual Path RNN with Attention} We extend Meta’s Dual Path RNN with the \(\texttt{DualPathRNN\_attention}\) class, introducing bidirectional LSTMs for better temporal coverage in both forward and backward directions. We also include an attention layer within each RNN block using the \(\texttt{SelfAttention}\) class. This allows the model to selectively focus on relevant signal features across time. The combination of bidirectional LSTMs and attention improves the model’s ability to capture complex temporal dependencies in neural data.

\paragraph{Final Convolutions} As in Meta’s model, our final stage includes two \(1 \times 1\) convolutions. The first projects to \(2D_2\) channels, and the second maps to \(F\) channels to match the dimensionality of the audio representation.


\section{Loss Functions}

While a regression loss such as mean squared error (MSE) could be used to train \(\mathbf{f}\), Defossez et al.\ argue that this is suboptimal for the task \cite{defossez2023decoding}. Instead, they propose a contrastive approach, specifically using CLIP Loss \cite{radford2021}, which encourages discriminative alignment between representations.

Given a segment of brain activity \(X\), a set of \(N\) candidate audio representations \(\{\bar{Y}_j\}_{j=1}^N\) is sampled, where \(\bar{Y}_N\) is the positive sample corresponding to \(X\), and \(\bar{Y}_1, \dots, \bar{Y}_{N-1}\) are negative samples from other (non-matching) audio segments. The model \(\mathbf{f}_{\text{clip}}\) maps \(X\) to a representation \(Z \in \mathbb{R}^{F \times T}\), which is then compared against each \(\bar{Y}_j\) using a dot product followed by softmax:

\begin{align}
    \hat{p}_j = \frac{\exp(\langle Z, \bar{Y}_j \rangle)}{\sum_{k=1}^N \exp(\langle Z, \bar{Y}_k \rangle)},
\end{align}

where \(\langle \cdot, \cdot \rangle\) denotes the inner product over the dimensions of \(Z\) and \(\bar{Y}_j\). The model is then trained to maximize the probability of the correct (positive) sample \(\bar{Y}_N\) using a standard cross-entropy loss between the true label \(p_j = \delta_{j,N}\) and the predicted probability \(\hat{p}_j\).


\section{Evaluation Metrics}

Evaluating model performance presents several challenges, primarily due to the scarcity of open datasets and the limited availability of reproducible code in existing studies. To address this, we establish specific criteria for our evaluation. As our model targets brain-computer interfaces where users think of words to control digital platforms, exact word-for-word accuracy isn't always necessary (i.e., capturing the intended meaning or command is often sufficient). Therefore, we introduce a dual-metric approach: traditional metrics such as Word Error Rate General (WER General) and Levenshtein Distance for exactness, and Word Error Rate Vocab (WER Vocab) for flexibility, ensuring that predicted words fall within the target vocabulary:

\begin{itemize}
     \item \textbf{Levenshtein Distance}: Measures the average number of single-character edits (insertions, deletions, substitutions) required to change predicted sentences into target sentences, normalized by the number of words in the target. For example, a value of 2.25 suggests that, on average, 2 to 3 edits are needed per word to correct the predictions.
     
    \item \textbf{WER General}: Calculated as the proportion of incorrectly identified words, where both position and order matter. For example, a WER General of 50\% means that 50\% of the words in the predictions differ from the target in exact sequence. We also report \textit{Accuracy General} as \textit{\(100 - \text{WER General}\)}.

    \item \textbf{WER Vocab}: Measures the proportion of words in the predictions that are not found in the target vocabulary, regardless of order or position. For example, a WER Vocab of 40\% means 40\% of the predicted words fall outside the target vocabulary. \textit{Accuracy Vocab} is calculated as \textit{\(100 - \text{WER Vocab}\)}, reflecting the percentage of predicted words found within the target vocabulary.
\end{itemize}


\section{Results and Discussion}

We use the work by Défosséz et al. (Meta) \cite{defossez2023decoding} as our baseline reference. This study is notable for (i) supporting cross-signal evaluation (MEG vs EEG), and (ii) providing open-source code and datasets for reproducibility. Our goal was to replicate their model using EEG data and extend it with targeted architectural modifications.

\subsection*{Baseline Replication}

Replicating Meta’s model on our EEG dataset yielded a vocabulary-level accuracy of 30.51\% on a subset of 20 subjects (compared to 25.75\% when using all 33). This aligns closely with Meta’s EEG results and serves as a solid foundation for our ablation studies.

\subsection*{Ablation and Architecture Modifications}

We evaluated multiple changes to the model pipeline (Table~\ref{tab:ablation-results}) to understand their contributions. Three modifications consistently improved accuracy: increasing the signal clamp value, using subject-specific layers, and integrating attention mechanisms. Conversely, removing dropout from the channel merger layer led to performance degradation.

\begin{table}[!htbp]
\centering
\small
\caption{Results of Architecture Variants and Key Design Changes}
\label{tab:ablation-results}
\begin{tabular}{*6c}
\toprule
Model 
& Clamp 
& Spatial
& Subject
& Dual Path 
& Accuracy \\
& Value
& Attention
& Layers
& RNN
& (Vocab) \\
\midrule
Meta (N=33, from paper)          & 20     & Dropout            & Subject layer      & LSTM          & 25.75\% \\
Meta (N=20, \textbf{our baseline}) & 20     & Dropout            & Subject layer      & LSTM          & 30.51\% \\
\midrule
Clamp ↑ (100)                     & 100    & default            & default            & default       & \textbf{34.98\%} \\
Clamp ↑ + Subject Embed       & 100    & No dropout         & Subject + Emb      & default       & \textbf{33.27\%} \\
Subject Embed Only            & default& default            & Subject + Emb      & default       & \textbf{38.41\%} \\
BiLSTM + Attn                & default& default            & default            & BiLSTM + Attn & 28.64\% \\
Subject Attn Layer           & default& default            & Subject + Attn     & default       & \textbf{30.66\%} \\
Subject-Specific Spatial Attn & default& Subject-specific   & default            & default       & \textbf{30.96\%} \\
\bottomrule
\end{tabular}
\end{table}

\paragraph{Clamp Values} Increasing the EEG signal clamp value from 20 to 100 resulted in a significant boost in decoding accuracy (from 30.51\% to 34.98\%). This likely enables the model to preserve more nuanced patterns in the EEG signal that may otherwise be truncated under lower thresholds.

\paragraph{Subject Embedding Layers} Introducing subject-specific embeddings substantially improved performance, especially when used independently (38.41\%) or combined with increased clamping (33.27\%). These embeddings act as learnable components that personalize the representation space for each subject, helping to mitigate inter-subject variability.

\paragraph{Spatial and Subject-Specific Attention} Adding subject-specific spatial attention (30.96\%) and a subject-specific attention layer (30.66\%) both yielded modest gains over the baseline (30.51\%). These mechanisms likely help the model focus on more informative channels and patterns relevant to each individual.

\paragraph{Dual-Path RNN with Attention} Replacing the single-directional LSTM with a bidirectional LSTM and post-layer attention led to degraded performance (28.64\%). This suggests that while theoretically more expressive, the additional complexity may introduce overfitting or interfere with alignment in this setting.


\section{Conclusions}

Decoding speech from EEG signals is a fundamentally difficult problem due to low signal-to-noise ratio and high inter-subject variability. Despite these challenges, EEG-based speech decoding is attractive for brain-computer interface (BCI) applications, given EEG's low cost and portability.

In this work, we build upon the state-of-the-art EEG decoder from Meta, introducing three targeted architectural modifications: subject-specific attention layers, personalized spatial attention, and a dual-path RNN with integrated attention. Two of these enhancements yielded measurable improvements in decoding performance, supporting the value of model personalization for this task.

Our results suggest that individualized attention mechanisms can help overcome some limitations of EEG by adapting to subject-specific neural patterns. This work lays the foundation for future improvements in non-invasive brain-to-speech decoding, with applications in accessibility and real-time BCI systems.


\section{Future Work}

While our results improve upon the Meta baseline, several directions remain for further research.

\paragraph{Scaling to a larger and more diverse subject pool}  Our experiments were conducted on 20 subjects (a subset of the 33 used by Meta) due to compute limitations. This limited sample may increase the risk of overfitting and reduce generalizability. Future work should include all available subjects and extend to new datasets to better capture inter-subject variability and validate the robustness of our findings.

\paragraph{Refining EEG pre-processing pipelines} Our results suggest that pre-processing significantly affects decoding performance. Future studies could explore varying the segment length (e.g., 5s, 10s, 15s) to provide richer context per example. Incorporating Independent Component Analysis (ICA) for artifact removal may also enhance signal quality, though it may interact with spatial attention mechanisms and requires careful tuning.

\paragraph{Integrating a language model for contextual decoding}  
EEG responses to speech are influenced by context, ambiguity, and partial word recognition. Incorporating a contextual language model (e.g., transformer-based decoder) could enable better inference of meaning from incomplete or ambiguous brain signals. This may bridge the gap between neural decoding and natural language understanding.

\paragraph{Prioritizing ethics and data privacy}  
As brain-computer interfaces become more powerful, ethical safeguards are essential. Future work should emphasize privacy, preserving model design, informed consent practices, and collaboration with ethicists and legal experts to ensure cognitive liberty and data security remain central to all deployments.


\section*{Acknowledgment}
We would like to thank Professor Bhiksha Raj of Carnegie Mellon University for his guidance and support throughout this project.

This work was conducted as part of the Carnegie Mellon University 11-785 Deep Learning course (Fall 2023):  
\href{https://deeplearning.cs.cmu.edu/F23/index.html}{https://deeplearning.cs.cmu.edu/F23/index.html}


\section*{Code}
The code for our experiments is available at:  
\url{https://github.com/kshapovalenko/DL-EEG-Speech-Decoder}


\bibliography{refs}


\end{document}